\documentclass[12pt]{article}
\makeatletter
\def\art{\@ifnextchar[{\eart}{\oart}}
\def\eart[#1]#2#3#4#5#6{{\rm #2}, {\em #3 \bf #4} {\rm (#6) #5} ({\em
#1})}
\def\hepart[#1]#2{{\rm #2, \em#1}}
\newcommand{\oart}[5]{{\rm #1}, {\em #2 \bf #3} {\rm (#5) #4}}
\newcounter{alphaequation}[equation]
\def\thealphaequation{\theequation\hbox to
0.6em{\hfil\alph{alphaequation}\hfil}}
\def\eqnsystem#1{
\def\@eqnnum{{\rm (\thealphaequation)}}
\def\@@eqncr{\let\@tempa\relax \ifcase\@eqcnt \def\@tempa{& & &} \or
  \def\@tempa{& &}\or \def\@tempa{&}\fi\@tempa
  \if@eqnsw\@eqnnum\refstepcounter{alphaequation}\fi
\global\@eqnswtrue\global\@eqcnt=0\cr}
\refstepcounter{equation} \let\@currentlabel\theequation \def\@tempb{#1}
\ifx\@tempb\empty\else\label{#1}\fi
\refstepcounter{alphaequation}
\let\@currentlabel\thealphaequation
\global\@eqnswtrue\global\@eqcnt=0 \tabskip\@centering\let\\=\@eqncr
$$\halign to \displaywidth\bgroup \@eqnsel\hskip\@centering
$\displaystyle\tabskip\z@{##}$&\global\@eqcnt\@ne
\hskip2\arraycolsep\hfil${##}$\hfil& \global\@eqcnt\tw@\hskip2\arraycolsep
$\displaystyle\tabskip\z@{##}$\hfil
\tabskip\@centering&\llap{##}\tabskip\z@\cr}

\def\endeqnsystem{\@@eqncr\egroup$$\global\@ignoretrue} \makeatother

\begin{document}

\def\lesssim{\mathrel{\mathpalette\vereq<}}
\def\gtrsim{\mathrel{\mathpalette\vereq>}}
\def\vereq#1#2{\lower3pt\vbox{\baselineskip1.5pt \lineskip1.5pt
\ialign{$\m@th#1\hfill##\hfil$\crcr#2\crcr\sim\crcr}}}
\makeatother

\newcommand{\NP}{Nucl. Phys.}
\newcommand{\PRL}{Phys. Rev. Lett.}
\newcommand{\PL}{Phys. Lett.}
\newcommand{\PR}{Phys. Rev.}

\newcommand{\rem}[1]{{\bf #1}}
\def\Red{}
\def\Black{}
\def\Blue{}

\renewcommand{\thefootnote}{\fnsymbol{footnote}}
\setcounter{footnote}{0}
\begin{titlepage}
\begin{center}

   \hfill    LBNL-42525\\
 \hfill    UCB-PTH-98/55\\
 \hfill       hep-ph/98\\

\vskip .5in

{\Large \bf \Red
Leading Order Textures for Lepton Mass Matrices
\footnote
{This work was supported in part by the U.S. 
Department of Energy under Contracts DE-AC03-76SF00098, in part by the 
National Science Foundation under grant PHY-95-14797. }
}\Black

\vskip .50in

Lawrence J. Hall and David Smith

\vskip 0.2in

{\em Department of Physics and\\ Theoretical Physics Group, Lawrence Berkeley National Laboratory\\
     University of California, Berkeley, California 94720}

\end{center}

\vskip .5in \Blue
\begin{abstract}
In theories with three light neutrinos, certain simplicity assumptions 
allow the construction of a complete list of leading order lepton mass 
matrices.  These matrices are consistent with $m_{\tau} \neq 0$, 
$\Delta m_{12}^2 \ll \Delta m_{23}^2$, $\theta_{23} \sim O(1)$ and
$\theta_{13} = 0$, as suggested by measurements of atmospheric and 
solar neutrino fluxes.  The list contains twelve generic cases: two 
have three degenerate neutrinos, eight have two neutrinos forming a 
Dirac state, and in only two cases is one neutrino much heavier than 
the other two.  For each of these twelve generic cases the possible 
forms for the perturbations which yield $m_{\mu}$ are given.  Ten 
special textures are also found.
\end{abstract}
\end{titlepage}

\renewcommand{\thepage}{\arabic{page}}
\setcounter{page}{1}
\renewcommand{\thefootnote}{\arabic{footnote}}
\setcounter{footnote}{0}

\section{Introduction} 
Over the last several decades, experiments have revealed a striking
generational pattern of quark and charged lepton masses and mixings.
In each charged sector there is a strong hierarchy of mass eigenvalues
between the three generations, $m_3 \gg m_2 \gg m_1$, and the three
angles describing mixing between the left-handed charge 2/3 quarks and
charge $-1/3$ quarks are all small. The origin of this pattern, and of
the precise values of the flavor observables, has been greatly
debated, with several diverse approaches and very many competing
theories.

Despite this diversity, a common theme can be identified: the fermion
masses are to be understood in an expansion, in which the leading
order term for each charged sector has the form
\begin{equation}
m^{(0)} \; = \; \pmatrix{0&0&0 \cr 0&0&0 \cr 0&0&A}.
\label{eq:usual}
\end{equation}
This gives the leading order results $m_3 \gg m_2 = m_1 =0$, and
vanishing mixing angles to the third generation. Indeed, for the
charged sectors it has seemed self evident that this {\em is} the
leading order structure, and the debate has centered on higher order
terms in the expansion.

The Super-Kamiokande measurements of the atmospheric neutrino fluxes \cite{superk}
cast considerable doubt on (\ref{eq:usual}) as the correct leading
order form, at least in the lepton sector. These measurements are of
great importance not just because they provide strong evidence for
neutrino masses: they may also fundamentally change our view 
of the pattern of flavor symmetry breaking.

The interpretation of this atmospheric $\nu$ flux data in terms of
neutrino oscillations implies a large mixing angle 
($\theta > 32^\circ $)
between $\nu_\mu$ and some other neutrino state, which could have
large $\nu_\tau$ or singlet neutrino components, but only a small
$\nu_e$ component. Is it possible to reconcile this observation with
the leading order form (\ref{eq:usual})? This issue is especially
important in unified theories, where relations are expected between 
the textures for the various charged sectors. We are aware of three
possible resolutions, each of which can be criticized:

\begin{itemize}
\item
In a three generation theory, with each generation containing a
right-handed neutrino, it is possible to write down textures for
charged leptons, ($m_E$), Dirac neutrino masses, ($m_{LR}$), and
right-handed Majorana masses, ($m_{RR}$), which all reduce to
(\ref{eq:usual}) at leading order, but which give a leading order form
to $m_{LL} = m_{LR} m_{RR}^{-1} m_{LR}^T$ which is very different from 
(\ref{eq:usual}), and has dominant terms giving $\theta_{\mu \tau}
\approx O(1)$. However, for this to happen the 23 and 33 entries of
$m_{LL}$ need to be comparable, and since they arise from different
terms in $m_{LR}$ and $m_{RR}$ the large value for $\theta_{\mu \tau}$
appears to be accidental.

\item
In a three generation theory, even if both $m_E$ and $m_{LL}$ have the
leading order form of (\ref{eq:usual}), the 23 entries may not be much
smaller than the 33 entries, generating a significant $\theta_{\mu
\tau}$ \cite{leontaris}. For example, in this conventional hierarchical scheme, the
ratios of eigenvalues in the charged and neutral sectors suggest
charged and neutral contributions to $\theta_{\mu \tau}$ of about
$14^\circ$ and $18^\circ$ respectively. Providing the relative
sign is such that these contributions add, large $\mu$-$\tau$
mixing can result. This is an important observation, because it shows
that the conventional picture, where all textures have the leading
order form of (\ref{eq:usual}), is not excluded by the Super-Kamiokande
data. However, the data does prefer an even bigger angle: the
conventional picture is now disfavored.\footnote{One might try to
argue that the conventional picture could even give $45^\circ$
mixing if the hierarchy between the two $\Delta m^2$ is reduced. This
is permissable if one of the solar neutrino experiments, or the standard solar
model, is incorrect \cite{bhssw}. However, in this case the 23 entry in $m_{LL}$ is
no longer small enough to be considered subleading.}

\item
In a theory with more than three light neutrino states it may be that 
(\ref{eq:usual}) gives the correct leading order neutrino masses
terms between the 3 left-handed states, but there is some additional
mass term coupling $\nu_\mu$ to a light singlet state leading to large
mixing between these states \cite{sterile}. Such schemes are certainly non-minimal,
and must answer three questions. Why is there a singlet state? Why is
it so light? Why is it coupled to $\nu_\mu$ rather than to $\nu_\tau$
or $\nu_e$?  Furthermore, during big bang nucleosynthesis the fourth state is
kept in thermal equillibrium by oscillations, and the resulting extra
contribution to the energy density is disfavored by observations which allow the primordial
abundances of D and $^4$He to be inferred \cite{shi}.

\end{itemize}

In view of these criticisms of the conventional leading order texture 
(\ref{eq:usual}), in this paper we study an alternative,
straightforward and direct 
interpretation of the data: in a three generation theory large
$\theta_{\mu \tau}$ arises because $m_{LL}$ and/or $m_E$ have a leading
order form which differs from (\ref{eq:usual}).
In the bulk of this paper, we perform an analysis to find all possible
leading order textures for $(m_E, m_{LL})$, subject to a simplicity
assumption, such that there is a hierarchy of neutrino mass
splittings: $\Delta m_{23}^2 \gg \Delta m_{12}^2$ as prefered by
atmospheric and solar neutrino data.\footnote{In \cite{alt1} a texture 
analysis is done to find the possible leading-order forms for 
$m_{LL}$, in the charge-diagonal basis, that have either maximal 
mixing for $\theta_{23}$ alone or maximal mixing for both 
$\theta_{23}$ and $\theta_{12}$.}

\section{Texture Analysis: Rules} 
In theories with three light neutrinos, the leading order, real, 
diagonal mass matrices consistent with 
$\Delta m_{12}^2 \ll \Delta m_{23}^2$ are
\begin{equation}
\overline{m}_{LL}^I \! =\! \pmatrix{0&0&0 \cr 0&0&0 \cr 0&0&\alpha} \hspace{0.1in}
\overline{m}_{LL}^{II}\! =\! \pmatrix{\alpha&0&0 \cr 0&\alpha&0 \cr 0&0&0} \hspace{0.1in}
\overline{m}_{LL}^{III}\! =\! \pmatrix{\alpha&0&0 \cr 0&\alpha&0 \cr 0&0&\beta} \hspace{0.1in}
\overline{m}_{LL}^{IV}\! =\! \pmatrix{\alpha&0&0 \cr 0&\alpha&0 \cr 0&0&\alpha} \hspace{0.1in}
\label{eq:nudiag}
\end{equation}
for the neutrinos, and
\begin{equation}
\overline{m}_E = \pmatrix{0&0&0 \cr 0&0&0 \cr 0&0&\gamma} 
\label{eq:chargediag}
\end{equation}
for the charged leptons.  In $\overline{m}_{LL}^{III}$, $\alpha$ and $\beta$ are of the same order but not equal.
The diagonal mass matrices are related to $m_{LL}$ and $m_E$, the mass matrices in the flavor 
basis, by unitary
transformations:
\begin{equation}
m_{LL} = V_{\nu}^* \overline{m}_{LL} V_{\nu}^{\dagger} \hspace{1.0in} 
m_{E} = V_{E_L} \overline{m}_{E} V_{E_R}^{\dagger}
\end{equation}
The leptonic mixing matrix $V= V_{E_L}^\dagger  V_{\nu}$ then relates the neutrino weak and mass
eigenstates according to 
\begin{equation}
\nu_{e_i} = V_{ij} \nu_{j}
\end{equation}
and can be parametrized by
\begin{equation}
V = R_{23}(\theta_{23})R_{13}(\theta_{13})\pmatrix{1&0&0 \cr 0&e^{i\beta} &0 \cr 0&0&1}R_{12}(\theta_{12})
\pmatrix{1&0&0 \cr 0&e^{i\alpha_1}&0 \cr 0&0&e^{i\alpha_2}}.
\label{eq:Vlong}
\end{equation}
If $\Delta m_{23}^2 > 2 \cdot 10^{-3}$ eV$^2$, then results from the Chooz experiment require 
$\theta_{13}< 13^\circ$ \cite{apollonio}.  In fact, even if $\Delta m_{23}^2 < 2 \cdot 10^{-3}$ eV$^2$, fits to the 
Super-Kamiokande atmospheric data (for $\Delta m_{12}^2 \ll \Delta m_{23}^2$) alone restrict
$\theta_{13}< 20^\circ$ \cite{barger}.  In light of these constraints we will assume that the leading order contribution to
$\theta_{13}$ vanishes, giving
\begin{equation}
V = R_{23}(\theta_{23})R_{12}(\theta_{12})\pmatrix{1&0&0 \cr 0&e^{i\alpha_1}&0 \cr 0&0&e^{i\alpha_2}},
\label{eq:V}
\end{equation}
with $\theta_{23}$ of order unity, as suggested by Super-Kamiokande results.

Our aim is to perform a systematic search for leading order leptonic mass matrices  $m_{LL}$ and $m_E$ that have 
the following features:
\begin{itemize}
\item Diagonalizing them gives ${\overline m}_E$ of (\ref{eq:chargediag}) for the charged leptons and
one of the ${\overline m}_{LL}$'s of (\ref{eq:nudiag}) for the
neutrinos. 
\item They produce a leptonic mixing matrix that can be paramatrized as in (\ref{eq:V}), with $\theta_{23} \sim 1$.
\item Their forms offer the hope of a simple explanation in terms of flavor symmetries.
\end{itemize}
Because we are particularly interested in leading order $m_{LL}$ and $m_E$ 
that can be simply understood using flavor
symmetries, we constrain their forms by allowing only the following exact relations between non-vanishing 
elements: 
\begin{itemize}
\item They may be equal up to a phase.
\item They may be related so as to give a vanishing determinant or sub-determinant.
\end{itemize}
The latter class of relations is allowed because, as discussed in 
\cite{{bhssw},{barbieri}, {alt2}},  vanishing determinants 
arise naturally when heavy particles are integrated out, as in the seesaw mechanism.  

As an illustration of how these rules are used, consider applying a (2-3) rotation, first on $\overline{m}_{LL}^I$, and 
second on $\overline{m}_{LL}^{II}$.  In the first case we get a neutrino mass matrix of the form 
\begin{equation}
m_{LL} = \pmatrix{0&0&0 \cr 0&{B^2 \over A}&B \cr 0&B&A}.
\end{equation}
Our rules allow this matrix because the relation among elements yields a vanishing
sub-determinant.  In the second case the transformation gives
\begin{equation}
m_{LL} = \pmatrix{{B^2 \over A} + A&0&0 \cr 0&{B^2 \over A}&B \cr 0&B&A}
\label{eq:illegal}
\end{equation}
(ignoring possible phases).  This matrix is not allowed because the relation between the 11, 22, and 33
entries is not essential for the vanishing of any determinant.  Cases such as (\ref{eq:illegal}) are not
excluded because it is impossible to attain them from a theory with a flavor symmetry.  Rather, they are 
excluded for reasons of simplicity:  in our judgement it is more difficult to construct such theories, compared
with theories for textures with all non-zero entries independent, equal 
up to a phase, or related to give a vanishing determinant. 

Given a pairing of leading order ($m_{LL}$, $m_E$) that satisfies our simplicity requirement, and that has mass eignenvalues 
consistent with (\ref{eq:nudiag}) and (\ref{eq:chargediag}), it is straightforward to determine whether or not $\theta_{23} \sim 1$
is satisfied.  Unfortunately, the remaining requirement, $\theta_{13} \sim 0$, is rendered meaningless by the leading order relation
$m_e = m_{\mu} = 0$.  This is easily seen by rotating the left-handed doublets to diagonalize $m_{LL}$, and then rotating the right-handed 
charged leptons to give
\begin{equation}
m_E = \pmatrix{0&0&0 \cr 0&0&0 \cr 0&0&A}, \hspace{0.1in} \pmatrix{0&0&0 \cr 0&0&B \cr 0&0&A},\hspace{0.1in} {\rm or}
\hspace{0.1in} \pmatrix{0&0&C \cr 0&0&B \cr 0&0&A}\hspace{0.15in} +\hspace{0.15in} {\rm perturbations}.
\label{eq:free}
\end{equation} 
We can diagonalize each leading order piece in (\ref{eq:free}) by applying (at most) a diagonal phase rotation followed by (1-2) and (2-3) rotations.
This indicates that, if we ignore the perturbations responsible for the muon mass, we are free to choose $\theta_{13} = 0$ for {\it any} 
leading order ($m_{LL}$, $m_E$) pairing.  

Although it is impossible to use the $\theta_{13} \sim 0$ requirement to restrict lepton mass matrices based on leading order 
considerations alone, it {\it is} true that it is easier for some ($m_{LL}$, $m_E$) pairings than it is for others to add perturbations 
that give $\theta_{13} \sim 0$.  As we will see, some pairings require special relations among the perturbations that seem
difficult to understand by symmetry considerations.  To exclude these cases we 
impose a final requirement on our leading order ($m_{LL}$, $m_E$) pairings:
\begin{itemize}
\item It must be possible to add to $m_{LL}$ and $m_E$ perturbations that establish $\theta_{13} \sim 0 $ and that satisfy the same simplicity requirements
already imposed on the leading order entries:  non-vanishing perturbations must be either independent, equal up to a phase, or related
in a way that gives a vanishing determinant.  We require that the perturbations in $m_E$ give $m_{\mu} \neq 0$ while
preserving $m_e = 0$.  For the case of three nearly degenerate neutrinos we require that the perturbations in $m_{LL}$
establish $\Delta m_{12}^2 \ll \Delta m_{23}^2$, 
and for the remaining cases, where $\Delta m_{23}^2 \neq 0$ is established at leading order,
we require that the perturbations in $m_{LL}$ lift the degeneracy between $\nu_1$ and $\nu_2$.
\end{itemize}
A simple example will clarify our motives for adding this requirement. 
Starting with the leading order textures
\begin{equation}
m_{LL} = \pmatrix{0&0&0 \cr 0&{B^2 \over A}&B \cr 0&B&A} \hspace{0.5in}
m_E = \pmatrix{0&0&0 \cr 0&0&E \cr 0&0&D},
\label{eq:easy}
\end{equation}
it is easy to find perturbations that satisfy our criteria.  For instance, if we add them according to 
\begin{equation}
m_{LL} = \pmatrix{0&0&0 \cr 0&{B^2 \over A}&B \cr 0&B&A+\epsilon_2} \hspace{0.5in}
m_E = \pmatrix{0&0&0 \cr 0&0&E \cr 0&\epsilon_1&D},
\label{eq:easypert}
\end{equation}    
then in the basis where $m_{LL}$ is diagonal we have
\begin{equation}
m_E = \pmatrix{0&0&0 \cr 0&{\epsilon_1}^\prime&E^\prime \cr 0&{\epsilon_2}^\prime&D^\prime},
\end{equation}
so that $\theta_{13} \sim 0$ is indeed satisfied, and the leading order matrices of (\ref{eq:easy}) are allowed.  
Things do not work as simply if we instead begin with the leading 
order pair
\begin{equation}
m_{LL} = \pmatrix{0&0&0 \cr 0&{B^2 \over A}&B \cr 0&B&A} \hspace{0.5in}
m_E = \pmatrix{0&0&F \cr 0&0&E \cr 0&0&D}.
\label{eq:hard}
\end{equation}
{\it After} rotating the lepton doublets to diagonalize $m_{LL}$ we need the form of $m_E$ (including perturbations responsible
for the muon mass) to have a perturbation only in the 32 entry:
\begin{equation}
m_E = \pmatrix{0&0&F \cr 0&0&E^\prime \cr 0& \epsilon & D^\prime}.
\label{eq:hardpert}
\end{equation}
Otherwise, after performing (1-2) and (2-3) rotations to diagonalize the leading order piece of $m_E$, we are still left with
an additional large (1-2) rotation required to diagonalize the perturbations, which induces a large
$\theta_{13}$.  One must therefore require that, in the flavor basis, the perturbations enter the charged lepton mass matrix
as in 
\begin{equation}
m_E = \pmatrix{0&0&F \cr 0&\epsilon B&E \cr 0&\epsilon A&D},
\label{eq:tune}
\end{equation} 
where $A$ and $B$ are the masses that appear in $m_{LL}$.  The non-trivial exact relation required between the perturbations
in $m_{E}$ and the leading order entries in $m_{LL}$ indicates that, for generic $A$ and $B$, the textures in (\ref{eq:hard}) do not
fulfill our criteria for leading order ($m_{LL}$,$m_E$).  Note, however, that for the special case $A = B$, the perturbations
in  (\ref{eq:tune}) are equal, so that the leading order pairing
\begin{equation}
m_{LL} = \pmatrix{0&0&0 \cr 0&A&A \cr 0&A&A} \hspace{0.5in}
m_E = \pmatrix{0&0&F \cr 0&0&E \cr 0&0&D}
\label{eq:special}
\end{equation} 
{\it is} allowed by our rules.

\section{Texture Analysis: Results}
The program of our analysis is as follows.  First, for each diagonal neutrino and charged lepton mass matrix 
of (\ref{eq:nudiag}) and (\ref{eq:chargediag}), we write down all possible forms for leading order $m_{LL}$ and $m_E$
in the flavor basis, consistent with our simplicity requirement restricting relations between non-vanishing
elements.  For each leading order ($m_{LL}$,$m_E$) pairing obtained in this way, we then determine whether there are perturbations
that satisfy the criteria described in the preceding paragraphs.

For example, for the case $\overline{m}_{LL} = \overline{m}_{LL}^I$ of equation (\ref{eq:nudiag}), the possible forms
for $m_{LL}$ in the flavor basis are
\begin{equation}
m_{LL} = \pmatrix{0&0&0 \cr0&0&0 \cr 0&0&A}, \hspace{0.2in}
m_{LL} = \pmatrix{0&0&0 \cr0&{B^2 \over A}&B \cr 0&B&A}, \hspace{0.1in} {\rm and} \hspace{0.1in}
m_{LL} = \pmatrix{{C^2 \over A}&{BC \over A}&C \cr {BC \over A}&{B^2 \over A}&B \cr C&B&A},
\label{eq:example}
\end{equation}
and all matrices obtained from these by permuting flavor basis indices.  Note that each relation among elements in these 
matrices leads to a vanishing sub-determinant, and is thus allowed.  These forms for $m_{LL}$ may be paired with either
\begin{equation}
m_{E} = \pmatrix{0&0&0 \cr0&0&0 \cr 0&0&D}, \hspace{0.2in}
m_{E} = \pmatrix{0&0&0 \cr0&0&E \cr 0&0&D}, \hspace{0.1in} {\rm or} \hspace{0.1in}
m_{E} = \pmatrix{0&0&F \cr 0&0&E \cr 0&0&D},
\label{eq:chargedexample}
\end{equation}
where only the left-handed charged leptons, and not necessarily the right-handed charged leptons, are in their
flavor basis\footnote{Because we consider forms for $m_{LL}$ obtained from those in (\ref{eq:example}) by permuting flavor basis indices, 
there is no need to do the same for $m_E$.  For example, we consider $m_{LL} = \pmatrix{{B^2 \over A}& 0& B
\cr 0&0&0 \cr B & 0 & A}$, but not $m_E = \pmatrix{0&0&B \cr 0&0&0 \cr 0&0&A}$.}.
Some ($m_{LL}$,$m_E$) pairings from (\ref{eq:example}) and (\ref{eq:chargedexample}) are immediately excluded because they
do not give $\theta_{23} \sim 1$,
\begin{equation}
m_{LL} = \pmatrix{0&0&0 \cr 0&0&0 \cr 0&0&A} \hspace{0.5in}
m_E = \pmatrix{0&0&0 \cr 0&0&0 \cr 0&0&D}
\label{eq:obvious}
\end{equation}  
being an obvious example. Other pairings, like that of (\ref{eq:hard}) for generic $A$ and $B$, are excluded because it not possible 
to add perturbtions that satisfy our requirements.
Some pairings that {\it do} work require exact relations between perturbations, as do the leading order textures in (\ref{eq:special}), 
while other pairings, such as the one in (\ref{eq:easy}), can accept independent perturbations.

Performing our analysis for each $\overline{m}_{LL}$ of equation (\ref{eq:nudiag}) leads to the pairings listed in
Tables 1-4.  Tables 1 and 2 list leading order ($m_{LL}$, $m_E$) pairings that can take perturbations with 
independent magnitudes; these twelve textures we call ``generic.'' Tables 3 and 4 contain pairings that instead 
require exact relations among perturbations, giving a further ten 
``special'' textures.  In Tables 1 and 2
we write the possible forms for $m_E$ as\footnote{More precisely, the various possible forms for $m_E$ can each be brought into one of these
three forms by appropriate rotations of the right-handed charged leptons. The perturbations are taken to 
have comparable magnitudes, but
in each matrix only $\epsilon_1$ need be non-zero.}
\begin{equation}
I \equiv \pmatrix{0&0&\epsilon_3 \cr0&\epsilon_1&\epsilon_2 \cr 0&0&D}, \hspace{0.2in}
II \equiv \pmatrix{0&0&\epsilon_2 \cr0&\epsilon_1&E \cr 0&0&D}, 
\hspace{0.1in} {\rm and} \hspace{0.1in}
III \equiv  \pmatrix{0&0&F \cr 0&0&E \cr 0&\epsilon_1&D}.
\label{eq:define}
\end{equation}
In Tables 3 and 4, we instead write $m_E$ explicitly and provide an example, for each pairing, of how
perturbations can be added to give acceptable masses and mixings.   Because for $m_{LL}$ there is often considerable 
freedom in how perturbations 
can be added, we show only leading order elements of $m_{LL}$, unless exact relations among these 
perturbations are required (as they are in the pairings with degenerate
neutrinos in Table 3).
\begin{table}
\centering
\begin{tabular}{|r|c|c|c|} \hline\hline
	&$m_{LL}$ & $m_E$ &$\overline{m}_{LL}$ \\ \hline
	1)&$\pmatrix{0&0&0 \cr 0&0&0 \cr 0&0&A}$ & $II$(U), 
	$III$(LA)&$\overline{m}_{LL}^I$ \\ &&& \\
	2)&$\pmatrix{0&0&0 \cr 0&{B^2 \over A}&B \cr 0&B&A}$ & $I$(U), 
	$II$(U)&$\overline{m}_{LL}^I$ \\ 
	\vspace{-.025in} &\vspace{-.025in}&\vspace{-.025in}&\vspace{-.025in} \\ \hline\hline 
	\vspace{-.025in} &\vspace{-.025in}&\vspace{-.025in}&\vspace{-.025in} \\ 
	3)&$\pmatrix{A&0&0 \cr 0&A&0 \cr 0&0&A}$ & $II$(U), 
	$III$(LA)&$\overline{m}_{LL}^{IV}$ \\&&&\\ 
	4)&$\pmatrix{0&A&0 \cr A&0&0 \cr 0&0&A}$ & $II$(LA), 
	$III$(LA)&$\overline{m}_{LL}^{IV}$\\ \hline\hline                       
\end{tabular}
\caption{The pairings of $m_{LL}$ and $m_E$ that can accept independent perturbations, and which have either 
a single massive Majorana neutrino or three degenerate neutrinos.  The 
matrices $I$, $II$, $III$, and $\overline{m}_{LL}^{I,II,III,IV}$ are as
defined in equations (\ref{eq:define}) and (\ref{eq:nudiag}); the meanings of U and LA are described above equation (\ref{eq:sa}).
$A$ and $B$ are independent complex parameters with comparable magnitudes.  
This list is complete, up to the freedom to relabel states in the flavor basis.}
\label{table:t1}
\end{table}

\begin{table}
\centering
\begin{tabular}{|r|c|c|c|} \hline\hline
	&$m_{LL}$ & $m_E$ &$\overline{m}_{LL}$ \\ \hline
	5)&$\pmatrix{A&0&0 \cr 0&A&0 \cr 0&0&0}$ & $II$(U), 
	$III$(LA)&$\overline{m}_{LL}^{II}$\\&&&\\
	6)&$\pmatrix{B&A&0 \cr A&-B&0 \cr 0&0&0}$ & $II$(LA), 
	$III$(LA)&$\overline{m}_{LL}^{II}$\\&&&\\
	7)&$\pmatrix{0&A&0 \cr A&0&0 \cr 0&0&0}$ & $II$(LA), 
	$III$(LA)&$\overline{m}_{LL}^{II}$\\&&&\\
	8)&$\pmatrix{0&A&B \cr A&0&0 \cr B&0&0}$ & $I$(LA), $II$(LA) 
	&$\overline{m}_{LL}^{II}$\\
	\vspace{-.025in} &\vspace{-.025in}&\vspace{-.025in}&\vspace{-.025in} \\ \hline\hline
	\vspace{-.025in} &\vspace{-.025in}&\vspace{-.025in}&\vspace{-.025in}\\
	9)&$\pmatrix{A&0&0 \cr 0&A&0 \cr 0&0&B}$ & $II$(U), 
	$III$(LA)&$\overline{m}_{LL}^{III}$\\&&&\\
	10)&$\pmatrix{B&A&0 \cr A&-B&0 \cr 0&0&C}$ & $II$(LA), 
	$III$(LA)&$\overline{m}_{LL}^{III}$\\&&&\\
	11)&$\pmatrix{0&A&0 \cr A&0&0 \cr 0&0&B}$ & $II$(LA), 
	$III$(LA)&$\overline{m}_{LL}^{III}$\\&&&\\
	12)&$\pmatrix{0&A&A \cr A&B&-B \cr A&-B&B}$ & $I$(LA) &$\overline{m}_{LL}^{III}$\\ \hline\hline
\end{tabular}
\caption{Same as Table \ref{table:t1}, but for $m_{LL}$'s that have 
$\nu_1$ and $\nu_2$ forming a pseudo-Dirac state.  
$A$, $B$, and $C$ are again independent complex parameters, except that in cases 6) and 10) 
there are certain phase relations.}
\end{table}
\begin{table}
\centering
\begin{tabular}{|r|c|c|c|} \hline\hline
	&$m_{LL}$ & $m_E$ &$\overline{m}_{LL}$ \\ \hline
	13)&$\pmatrix{0&0&0 \cr 0&A&A \cr 0&A &A}$ & $\pmatrix{0&0&F \cr 0&\epsilon_1&E \cr 0&\epsilon_1
	&D}$(LA) & $\overline{m}_{LL}^{I}$\\&&&\\
	14)&$\pmatrix{A&0&A \cr 0&0&0 \cr A&0&A}$ & 
	$\pmatrix{0&\epsilon_1&\epsilon_2 \cr 0&0&E \cr 0&\epsilon_1&D}$(LA) &$\overline{m}_{LL}^{I}$\\&&&\\
	15)&$\pmatrix{{B^2 \over A}&{B^2 \over A}&B \cr {B^2 \over A}&{B^2 \over A}&
	B \cr B&B&A}$ & $\pmatrix{0&\epsilon_1&\epsilon_2 \cr 0&\epsilon_1&\epsilon_2 \cr 
	0&0&D}$(U)&$\overline{m}_{LL}^{I}$\\&&&\\
	16)&$\pmatrix{A&A&A \cr A&A&A \cr 
	A&A&A}$ & $\pmatrix{0&\epsilon_1&\epsilon_2 \cr 0&\epsilon_1&E \cr 0&\epsilon_1&D}$(U),
	$\pmatrix{0&\epsilon_1&F \cr 0&\epsilon_1&E \cr 0&\epsilon_1&D}$(LA) 
	&$\overline{m}_{LL}^{I}$\\
	\vspace{-.025in} &\vspace{-.025in}&\vspace{-.025in}&\vspace{-.025in} \\ \hline\hline
	\vspace{-.025in} &\vspace{-.025in}&\vspace{-.025in}&\vspace{-.025in}\\
	17)&$\pmatrix{A+\epsilon_1&0&0 \cr 0&A&\epsilon_1 \cr 0&\epsilon_1&A}$ & $\pmatrix{0&0&\epsilon_4 \cr 0&\epsilon_2&\epsilon_3
	\cr 0&0&D}$(U) & $\overline{m}_{LL}^{IV}$\\&&&\\
	18)&$\pmatrix{A+\epsilon_1&0&0 \cr 0&\epsilon_1&A \cr 0&A&\epsilon_1}$ & $\pmatrix{0&0&\epsilon_4 \cr 0&\epsilon_2&\epsilon_3
	\cr 0&0&D}$(U) &$\overline{m}_{LL}^{IV}$\\ \hline\hline
\end{tabular}
\caption{The pairings of $m_{LL}$ and $m_E$ that require exact relations among perturbations, and which have either a 
single massive Majorana neutrino or three degenerate neutrinos.  $A$ - $F$ are independent complex parameters with comparable
magnitudes, as are the perturbations $\epsilon_1$, $\epsilon_2$, 
$\epsilon_3$, and $\epsilon_4$.  This is a complete list of leading 
order textures, up to the 
freedom to relabel states in the flavor basis.  The perturbations are shown simply to
illustrate, for each leading order pairing, how they can be included in a way consistent with our requirements.}
\end{table}

\begin{table}
\centering
\begin{tabular}{|r|c|c|c|} \hline\hline
	&$m_{LL}$ & $m_E$ &$\overline{m}_{LL}$ \\ \hline
	19)&$\pmatrix{0&A&A \cr A&0&0 \cr A&0&0}$ & $\pmatrix{0&0&F \cr 0&\epsilon_1&E \cr 0&-\epsilon_1&D}$(LA)
	&$\overline{m}_{LL}^{II}$\\&&&\\
	20)&$\pmatrix{0&A&0 \cr A&0&A \cr 0&A&0}$ & $\pmatrix{0&\epsilon_1&\epsilon_2 \cr 0&0&E \cr 0&-\epsilon_1&D}$(LA)
	&$\overline{m}_{LL}^{II}$\\
	\vspace{-.025in} &\vspace{-.025in}&\vspace{-.025in}&\vspace{-.025in} \\ \hline\hline
	\vspace{-.025in} &\vspace{-.025in}&\vspace{-.025in}&\vspace{-.025in}\\
	21)&$\pmatrix{0&A&A \cr A&B&-B \cr A&-B&B}$ & $\pmatrix{0&0&F \cr 0&\epsilon_1&E \cr 0&-\epsilon_1&D}$(LA)
	&$\overline{m}_{LL}^{III}$\\&&&\\
	22)&$\pmatrix{B&A&-B \cr A&0&A \cr -B&A&B}$ & $\pmatrix{0&\epsilon_1&\epsilon_2 \cr 0&0&E \cr 0&-\epsilon_1&D}$(LA)
	&$\overline{m}_{LL}^{III}$\\ \hline\hline
\end{tabular}
\caption{Same as Table 3, but for $m_{LL}$'s that have 
$\nu_1$ and $\nu_2$ forming a pseudo-Dirac state.}  
\label{table:t2}
\end{table}
\smallskip

Some of the pairings of leading order $m_{LL}$ and $m_E$ in these tables lead to equivalent 
physics;  for instance, the masses
\begin{equation}
m_{LL} = \pmatrix{0&0&0 \cr 0&0&0 \cr 0&0&A}
\hspace{0.5in}
m_E = \pmatrix{0&0&0 \cr 0&0&E \cr 0&0&D}
\label{eq:physeq1}
\end{equation}
are related by a simultaneous (2-3) rotation on both the charged leptons
and neutrinos to the combination
\begin{equation}
m_{LL} = \pmatrix{0&0&0 \cr 0&{B^2 \over A}&B \cr 0&B&A}
\hspace{0.5in}
m_E = \pmatrix{0&0&0 \cr 0&0&E^{\prime} \cr 0&0&D^{\prime}}.
\label{eq:physeq2}
\end{equation}
As a consequence (\ref{eq:physeq1}) and (\ref{eq:physeq2}) give the same 
form for the leptonic mixing matrix and are thus physically indistinguishable.
For our purposes, (\ref{eq:physeq1}) and (\ref{eq:physeq2}) represent
distinct cases because theories that predict the mass matrices of
(\ref{eq:physeq1}) in the flavor basis will be different from those that predict
the mass matrices of (\ref{eq:physeq2}).  In other words, the apparent 
redundancy among some of the pairings of Tables 1 - 4 arises because our rules were
implemented with model-building purposes in mind.

In fact, some of the leading order ($m_{LL}$,{$m_E$}) combinations that at first sight seem to lead to the same physics
emerge as less similar once we consider the effects of perturbations.  For example, due to the degeneracy
of $\nu_1$ and $\nu_2$, we can find a simultaneous transformation that brings the matrices
\begin{equation}
m_{LL} = \pmatrix{0&0&0 \cr 0&0&0 \cr 0&0&A}
\hspace{0.5in}
m_E = \pmatrix{0&0&F \cr 0&0&E \cr 0&0&D}
\label{eq:three}
\end{equation}
into the forms 
\begin{equation}
m_{LL} = \pmatrix{0&0&0 \cr 0&0&0 \cr 0&0&A}
\hspace{0.5in}
m_E = \pmatrix{0&0&0 \cr 0&0&E \cr 0&0&D}.
\label{eq:two}
\end{equation}
However, we know that this degeneracy is lifted by perturbations in $m_{LL}$, so that the similarity between (\ref{eq:three})
and (\ref{eq:two}) is somewhat artificial.  For the matrices in (\ref{eq:two}), the perturbations alone determine 
$\theta_{12}$, which can turn out to be arbitrarily large or small.  For the matrices in (\ref{eq:three}), on the
other hand, we generally expect $\theta_{12} \sim 1$, barring an unlikely near-cancellation between
the (1-2) rotation induced by the perturbations in $m_{LL}$ and the (1-2) rotation required to diagonalize  
$m_E$ at leading order.  Note that conversely, the physical equivalence we identified between
(\ref{eq:physeq1}) and (\ref{eq:physeq2}) does not rely on the degeneracy between $\nu_1$ and $\nu_2$, so that these matrices
are on similar footing with regard to their response to perturbations.

For each pairing in Tables 1 - 4 we
have identified whether, as in (\ref{eq:two}), the size of $\theta_{12}$ is fixed entirely by perturbations, so that no indication is given 
regarding which solutions to the solar neutrino problem
are favored (denoted by ``U''), or whether, as in (\ref{eq:three}), we typically have $\theta_{12} \sim 1$, so that large angle solutions 
are favored (denoted by ``LA'').
Although we have not listed them explicitly, there are in fact pairings of leading order $m_{LL}$ and $m_E$ that 
require small angle MSW solutions to 
the solar neutrino problem \cite{msw}.  For example, the pairing
\begin{equation}
m_{LL} = \pmatrix{0&A&0 \cr A&0&0 \cr 0&0&0}
\hspace{0.5in}
m_E = \pmatrix{0&0&E \cr 0&0&E \cr 0&0&D}
\label{eq:sa}
\end{equation}
is a special case of one of the combinations in 7) from Table 2\footnote{We regard mass matrices obtained  
by setting, for example, $A = B$ in matrices from
Tables 1 - 4 as special cases, and do not list 
them independently, even though matrices like
$m_E = \pmatrix{0&0&0 \cr 0&0&E \cr 0&0&D}$ and $m_E = \pmatrix{0&0&0 \cr 0&0&D \cr 0&0&D}$ are quite different from a model builder's
perspective, since different symmetries would be required to motivate them.}, and can {\it only} give
a small angle MSW solution.

Up to this point we have said nothing about complex phases in our matrices.  To ensure that the leading order relation $\Delta m^2_{12} = 0$ holds,
we must require that in the $m_{LL}$'s of pairings 6) and 10), the $A$'s and $B$'s share the same phase, up to the freedom to 
send $\nu_i \rightarrow e^{i \alpha_i} \nu_i$.  This means, for instance, that the $m_{LL}$ in 6) actually stands for
\begin{equation}
m_{LL} = \pmatrix{Be^{2i\alpha} &Ae^{i(\alpha+\beta)}&0 \cr Ae^{i(\alpha+\beta)}&-Be^{2i\beta}&0 \cr 0&0&0},
\end{equation}
with $\alpha$ and $\beta$ arbitrary and $A$ and $B$ real.  In all other pairings, the phases of $A$ - $F$ and the various $\epsilon$'s 
are independent\footnote{Moreover, the freedom to send $\nu_i \rightarrow e^{i \alpha_i} \nu_i$ allows the two $A$'s in
the $m_{LL}$ of 5), for instance, to have different phases.}.

\section{Some Special Textures}
In this section we discuss specific features of some of the more interesting 
pairings in Tables 1 - 4.\\
\underline{$\theta_{23}={\pi \over 4}$}\\
One simple possibility, consistent with data from Super-Kamiokande, is that the leading order lepton 
masses give precisely $\theta_{23}={\pi \over 4}$.  For
a neutrino mass matrix that requires no (2-3) rotation, the charged lepton mass matrix
\begin{equation}
m_E = \pmatrix{0&0&0 \cr 0&0&D \cr 0&0&D}
\end{equation}
gives maximal mixing.  Conversely, if the charge lepton mass matrix assumes the form
\begin{equation}
m_E = \pmatrix{0&0&0 \cr 0&0&0 \cr 0&0&D},
\end{equation}    
then the forms for $m_{LL}$ from Tables 1 and 2 that give $\theta_{23}={\pi \over 4}$ are
\begin{equation}
\hspace{-0.05in}
m_{LL} = \pmatrix{0&A&A \cr A&0&0 \cr A&0&0}, \hspace{0.17in}
m_{LL} = \pmatrix{0&0&0 \cr 0&A&-A \cr 0&-A&A}, \hspace{0.07in}{\rm and} \hspace{0.07in}
m_{LL} = \pmatrix{0&A&A \cr A&B&-B \cr A&-B&B}.
\end{equation}\\
Other pairings that give maximal mixing require exact relations among perturbations, and can be found in
17) and 18) of Table 3.\\
\underline{\it{Neutrinos as hot dark matter}}\\
If there exist three light neutrinos whose splittings obey $\Delta m_{12}^2 \ll 
\Delta m_{23}^2 \sim 10^{-3} eV^2$, then for neutrino masses to be cosmologically significant
requires a high degree of degeneracy.  Furthermore, there is a bound from neutrinoless double $\beta$ 
decay experiments that, in the basis where the charged lepton masses are diagonal,
${m_{LL}}_{ee}<.5eV$ \cite{baudis}.  Lowest-order mass matrices that give degenerate neutrinos {\it and}
${m_{LL}}_{ee} = 0$ are thus of special interest, as they evade this experimental constraint and
allow the neutrino mass scale to be cosmologically relevant.  We find two combinations of $m_{LL}$
and $m_E$ that satisfy these criteria:
\begin{equation}
m_{LL} = \pmatrix{A&0&0 \cr 0&A&0 \cr 0&0&A} \hspace{0.5in}
m_E = \pmatrix{0&0&iE \cr 0&0&E \cr 0&\epsilon&D}
\label{eq:hdm1}
\end{equation}
and
\begin{equation}
m_{LL} = \pmatrix{0&A&0 \cr A&0&0 \cr 0&0&A} \hspace{0.5in}
m_E = \pmatrix{0&0&0 \cr 0&\epsilon_2&E \cr 0&\epsilon_1&D}.
\label{eq:hdm2}
\end{equation} 
We include perturbations in $m_E$ because without them, the element 
${m_{LL}}_{ee}$ is not defined.\\
\underline{\it{Zeroth order lepton mass matrices}}\\
We call the contributions to $m_{LL}$ and $m_E$ that survive in the limit of unbroken flavor
symmetry ``zeroth order'' masses.  Consider the case of an abelian 
flavor symmetry, and suppose that both $m_{LL}$ and $m_E$ have non-vanishing
elements at zeroth order.  If in addition the zeroth order form of $m_E$ in the flavor basis is 
invariant under
$\nu_i \leftrightarrow \nu_j$, then it follows that $\nu_i$ and $\nu_j$ are not
distinguished by the flavor symmetry: if $\nu_{i}\nu_{i}h_{u}$ is an 
allowed operator, then so are $\nu_{i}\nu_{j}h_{u}$ and 
$\nu_{j}\nu_{j}h_{u}$.  As a consequence it must be true that $m_{LL}$ as well
is invariant under $\nu_i \leftrightarrow \nu_j$, and moreover that the (i-j) space of $m_{LL}$ must have either
all entries zero, or all entries non-zero.  Following this reasoning, we 
find that, for abelian flavor symmetries, the 
only pairings from Tables 1 - 4 that are candidate zeroth order mass matrices are
\begin{equation}  
m_{LL} = \pmatrix{0&A&B \cr A&0&0 \cr B&0&0} \hspace{0.5in}
m_{E} = \pmatrix{0&0&0 \cr 0&0&E \cr 0&0&D},
\label{eq:zerotha}
\end{equation}
and
\begin{equation}
m_{LL} = \pmatrix{0&0&0 \cr 0 & {B^2 \over A} &B \cr 0&B&A} \hspace{0.5in}
m_{E} = \pmatrix{0&0&0 \cr 0&0&E \cr 0&0&D}.
\label{eq:zerothb}
\end{equation}
Simple seesaw-based models for the combinations in (\ref{eq:zerotha}) and (\ref{eq:zerothb}) were described in \cite{barbieri};
several other models have been based on the $m_{{LL}}$ of (\ref{eq:zerothb}) 
\cite{berezhiani}.\\
\underline{\it{Democratic mass matrices}}\\
The pairing
\begin{equation}
m_{LL} = \pmatrix{0&0&0 \cr 0 & 0 &0 \cr 0&0&A} \hspace{0.5in}
m_{E} = \pmatrix{0&0&D \cr 0&0&D \cr 0&0&D},
\end{equation}
which is a special case ($D=E=F$) of one of the combinations in 1), results from 
a leading order democratic form for the charged matrix 
\cite{democratic}.  A democratic 
form for the neutrino mass matrix is generically excluded, since it 
gives too large a value for $\theta_{13}$.  However, it is allowed 
with special perturbations, as shown in pairing 16).  
\section{Limitations of Texture Analysis}
Some of the requirements imposed on $m_{LL}$ and $m_E$ in section 2 were motivated by a desire to
concentrate on mass matrices that could result most easily from flavor symmetries.  One may wonder what
we have missed in this regard - are there forms for $m_{LL}$ and $m_E$ that violate our rules but that
nevertheless are plausable as consequences of flavor symmetry?  

One reason that examples of such matrices can in fact be found is that if $m_{LL}$ arises by
the seesaw mechanism \cite{seesaw}, then our rules should really be applied to $m_{RR}$ and $m_{LR}$, and
$m_{LL}$ should be {\it derived} from these matrices according to 
\begin{equation}
m_{LL} = m_{LR}m_{RR}^{-1}m_{LR}^T.
\end{equation}
For example, the matrices
\begin{equation}
m_{RR} = \pmatrix{0&-A&A \cr -A&A&0 \cr A&0&0} \hspace{0.3in} {\rm and} \hspace{0.3in}
m_{LR} = \pmatrix{0&0&A \cr 0&A&A \cr A&A&A}
\end{equation}
are certainly consistent with our rules, while the resulting mass matrix for the light neutrinos,
\begin{equation}
m_{LL} = A \pmatrix{1&2&3 \cr 2&4&5 \cr 3&5&6},
\end{equation}
clearly is not.  Examples like this are not difficult to find, but it does seem to be
true that in most simple cases, if $m_{RR}$ and $m_{LR}$ satisfy our rules, then so does $m_{LL}$.

Another limitation of our approach is that our rules prohibit matrix  elements from being related by
factors of $1 \over 2$, $1 \over 3$, etc., that could arise from Clebsch coefficients associated with 
the flavor
group.  For example, in the basis 
where the charged lepton masses are diagonal, the form
\begin{equation}
m_{LL} =A \pmatrix{0&{1 \over {\sqrt 2}}&{1 \over {\sqrt 2}} \cr {1 \over {\sqrt 2}}&{1 \over 2}&
-{1 \over 2} \cr {1 \over {\sqrt 2}}&-{1 \over 2}&{1 \over 2}}
\label{eq:gg}
\end{equation}
allows (for large enough $A$) neutrinos to compose a significant fraction of the dark 
matter in the universe without violating double $\beta$ decay constraints \cite{georgi}. 
This texture corresponds to the pairing 12) of Table 2, 
with $B$ chosen to be $A \over {\sqrt 2}$, in violation of our 
rules.  On the other hand, the procedure we have used {\it has} 
given us a pairing of $m_{LL}$ and $m_E$ that leads precisely to the same physics 
as does (\ref{eq:gg}).  In particular, (\ref{eq:hdm2}) for the special 
case $D = E$ is identical to (\ref{eq:gg})
as far as the physics is concerned, and we expect that these forms may be easier to
motivate with flavor symmetries.  
    
\section{Conclusions}
The Super-Kamiokande data on atmospheric neutrino fluxes suggests 
that the leading-order fermion mass matrices may not have the 
conventional form of (\ref{eq:usual}), at least in the lepton sector.  
What leading order forms for lepton mass matrices are suggested by 
atmospheric and solar neutrino oscillations?  We have derived the 
complete set of leading order ($m_{LL}$,$m_E$) pairings consistent with
$\Delta m_{12}^2 \ll \Delta m_{23}^2$, $\theta_{13} \sim 0$, and $\theta_{23} \sim 1$, 
subject to a simplicity requirement: non-zero entries of a matrix may 
be equal up to a phase, or may have a precise relationship which 
leads to a vanishing determinant, otherwise they are independent. This simplicity requirement, motivated
by an interest in textures that follow most easily from flavor symmetries, 
reduces an infinitely large class of matrices to the ($m_{LL}$,$m_E$) combinations listed in 
Tables 1 - 4.  

These combinations are divided into twelve generic cases and ten 
special cases, according to whether the perturbations involve exact 
relations.  For the twelve generic cases we also give the possible 
forms for the perurbations responsible for the muon mass.  The diverse pairings we have derived lead to a variety of physics.  
Some give degenerate neutrinos and thus leave considerable freedom in the overall mass scale, 
while others, with hierarchical masses, fix the mass scale of the heaviest neutrino at $\sim 3 \times
10^{-2}$eV, according to Super-Kamiokande results.
The various pairings also give different predictions for $\theta_{12}$, and hence require different
resolutions to the solar neutrino problem. Certainly each of our mass matrices is incomplete, 
because only by specifying all perturbations can the physics be fully established, 
yet in our view the approach we have taken offers a simple starting point for considering what 
mass matrices to aim for in constructing realistic theories of lepton 
masses.

\end{document}